\documentclass[preprint,authoryear,12pt]{elsarticle}

\makeatletter
\def\ps@pprintTitle{%
 \let\@oddhead\@empty
 \let\@evenhead\@empty
 \def\@oddfoot{}%
 \let\@evenfoot\@oddfoot}
\makeatother

\usepackage{epsfig}

\usepackage{amssymb}

\usepackage[a4paper=true,%
breaklinks=true,%
colorlinks=true,%
pdfauthor={First Author et al.},%
pdftitle={Validation of the in-flight calibration procedures for the MICROSCOPE space mission}%
]{hyperref}

\begin{document}

\begin{frontmatter}


\title{Validation of the in-flight calibration procedures for the MICROSCOPE space mission}

\author{{\'{E}}milie Hardy\corref{cor}}
\address{ONERA, 29 avenue de la Division Leclerc, F-92322 Ch{\^{a}}tillon, France}
\cortext[cor]{Corresponding author}
\ead{emilie.hardy@onera.fr}

\author{Agn{\`{e}}s Levy, Manuel Rodrigues and Pierre Touboul}
\address{ONERA, 29 avenue de la Division Leclerc, F-92322 Ch{\^{a}}tillon, France}
\ead{agnes.levy@onera.fr, manuel.rodrigues@onera.fr, pierre.touboul@onera.fr}

\author{Gilles M{\'{e}}tris}
\address{G{\'{e}}oazur, Universit{\'{e}} de Nice Sophia-Antipolis, Centre National de la Recherche Scientifique (UMR 7329), Observatoire de la C{\^{o}}te d'Azur, 250 avenue Albert Einstein, 06560 Valbonne, France}
\ead{gilles.metris@oca.eu}

\begin{abstract}
The MICROSCOPE space mission aims to test the Equivalence Principle with an accuracy of $10^{-15}$. The drag-free micro-satellite will orbit around the Earth and embark a differential electrostatic accelerometer including two cylindrical test masses submitted to the same gravitational field and made of different materials. The experience consists in testing the equality of the electrostatic acceleration applied to the masses to maintain them relatively motionless. The accuracy of the measurements exploited for the test of the Equivalence Principle is limited by our a priori knowledge of several physical parameters of the instrument. These parameters are partially estimated on-ground, but with an insufficient accuracy, and an in-orbit calibration is therefore required to correct the measurements. The calibration procedures have been defined and their analytical performances have been evaluated. In addition, a simulator software including the dynamics model of the instrument, the satellite drag-free system and the perturbing environment has been developed to numerically validate the analytical results. After an overall presentation of the MICROSCOPE mission, this paper will describe the calibration procedures and focus on the simulator. Such an in-flight calibration is mandatory for similar space missions taking advantage of a drag-free system.
\end{abstract}


\end{frontmatter}

\parindent=0.5 cm

\section{Introduction}

MICROSCOPE is a fundamental physics space mission that aims to test the Equivalence Principle (EP) with a precision never achieved before. The Equivalence Principle is the basis of Einstein's theory of General Relativity, which has never been disproved by the observations since its elaboration one century ago.

However, General Relativity does not allow the unification of the four fundamental interactions. Electromagnetism, as well as the weak and strong interactions are described by Quantum Mechanics, while Gravitation is described by General Relativity. Einstein's theory is incompatible with the formalism that describes the three other fundamental interactions. To unify the gravitation with the other interactions, General Relativity therefore needs to be reconsidered. New theories are thus under development to allow the great unification, for example the string theory or the supersymmetry. Some of these theories predict a violation of the Equivalence Principle at levels that have not been tested yet \citep{Damour-ViolationEquivalencePrinciple}.

Testing the Equivalence Principle beyond this limit is therefore an important verification of General Relativity, and a way to constrain the newly developed gravitational theories. One of the EP simplest manifestations is the Universality of Free Fall, which states that inertial and gravitational masses are equivalent. Thus, the trajectory of a free falling object (only submitted to gravity) is independent of its internal structure and composition. The Universality of Free Fall has been tested throughout the centuries with an improving accuracy. Important results have been obtained thanks to the Lunar Laser Ranging method, even though the interpretation is limited by the imprecision on the compositions of Earth and Moon \citep{Williams-LunarLaserRanging}. Experiences using rotating torsion balance have led to a record accuracy of a few $10^{-13}$ \citep{Schlamminger-TorsionBalance}, but these on-ground experiments are highly disturbed by environmental instabilities. Being performed in space, the MICROSCOPE experiment will be able to reach a new step of accuracy of $10^{-15}$ and will open the way to even more ambitious space missions.

	The test is based on the precise measurement delivered by a dedicated differential accelerometer accommodated on board a drag-free micro-satellite orbiting around the Earth. The EP test accuracy is limited among other things by the a priori knowledge of the instrument's physical parameters. The on-ground evaluation of these instrumental parameters is not precise enough and an in-orbit calibration is required to finely determine them and correct the measurements from their effects. Contrary to other missions with ultra-sensitive accelerometers like CHAMP or GRACE \citep{VanHelleputte-CHAMPandGRACE}, this calibration can take advantage of the satellite drag-free system.
	
After the overall presentation of the MICROSCOPE mission and its payload, the mathematical expression of the accelerometer differential measurement is detailed exhibiting the perturbing terms. The in-orbit calibration procedures are then depicted and the expression of the estimates and the corrections provided. We will then focus on the software simulator developed to validate numerically these procedures.

\section{Overview of the MICROSCOPE project}
\label{sec:2}

\subsection{The MICROSCOPE experiment}
\label{ssec:2.1}

\begin{figure}
  \centering
  \includegraphics[width=0.7\textwidth,clip]{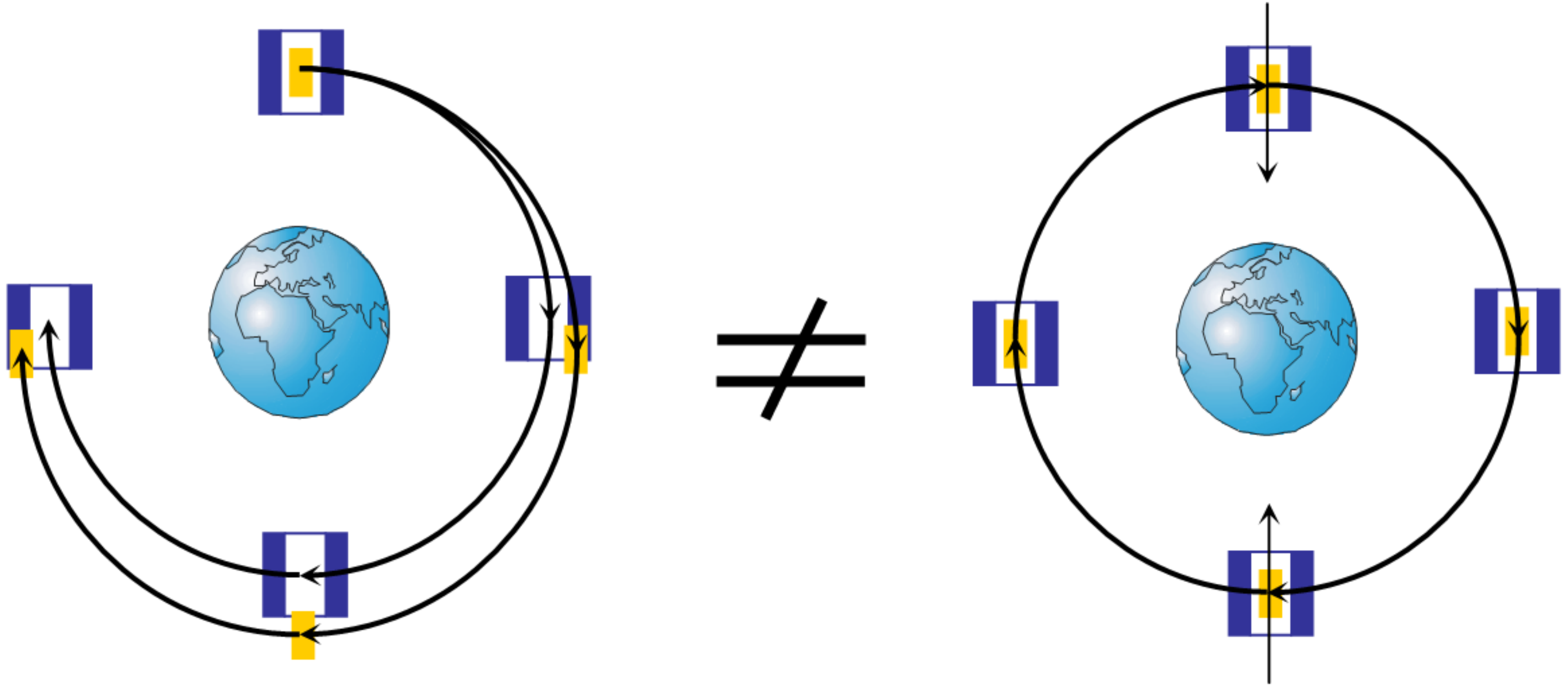}
  \caption{On the left, the satellite orbits around the Earth with two concentric test masses (yellow and blue) of different composition. A difference in their trajectories indicates a violation of the Equivalence principle. For the MICROSCOPE experiment, on the right, the measurement is not the difference of trajectories, but the difference in the forces applied to maintain the masses relatively motionless.}
  \label{fig:orbite-terrestre}
\end{figure}

MICROSCOPE will test the Universality of Free Fall, an expression of the Equivalence Principle, by comparing the accelerations of two masses constituted by different materials on board a micro-satellite orbiting around the Earth. The cylindrical and concentric masses are maintained electrostatically levitated and are servo-controlled to follow the same orbit with a precision better than $10^{-11} \, \mbox{m}$, so that they will undergo the same gravity field if they are perfectly centered (see figure \ref{fig:orbite-terrestre}). This is made possible by the electrostatic actuation which forces the masses to remain relatively motionless. If the perturbations are well enough controlled, a difference measured between the electrostatic forces applied to the masses will indicate a violation of the Equivalence Principle.

The MICROSCOPE experiment takes place on board a micro-satellite developed by the Centre National d'{\'{E}}tudes Spatiales (CNES) within its MYRIAD micro-satellite program. The satellite is equipped with a drag-free system that compensates for the surface forces, such as the Sun and Earth radiation pressures and the residual atmospheric drag. Thus, the satellite motion is finely servo-controlled along the orbit and in attitude. The command of the drag-free system is the data provided by the two inertial sensors composing the differential accelerometer \citep{Touboul-2001}.

The orbit will be quasi-circular, with an eccentricity lower than $5 \times 10^{-3}$, and heliosynchronous, at an altitude around $720 \, \mbox{km}$. This altitude is a compromise between the intensity of the Earth gravitational signal and the residual atmospheric drag to be compensated. It provides an orbital period of about $5950 \, \mbox{s}$. The heliosynchronous orbit allows fixed and efficient solar panels. Moreover, this helps for the thermal stability of the experiment because of the continuous exposition of the same external side of the satellite to the solar radiations. The instrument's thermal control can then be passive and compatible with the accommodation constraints of the micro-satellite.  

One of the most important advantages of this space experiment is the time span of the measurement which is not limited by the free-fall. In the MICROSCOPE experiment, the measurement duration of our test will be superior to 20 orbits in order to integrate the signal and reduce the noise. Moreover, the electric, magnetic, thermal, gravitational and vibrational environment is very steady and can be controlled. The use of the Earth instead of the Sun as the source of the gravity field allows the increase of the signal to be detected by more than three orders of magnitude. In addition, the frequency and phase of the signal are very well defined, depending on the Earth gravity signal projected in the instrument reference frame. The satellite pointing can be either inertial or spinning. In the first case, the effects of the centrifugal acceleration perturbations are limited and the Earth gravity field is modulated by the orbital frequency. The Equivalence Principle frequency is therefore $f_{EP_i} = f_{orb} = 0,17 \, \mbox{mHz}$. In the second case, the signal frequency $f_{EP_s}$ is the sum of the spin and the orbital frequencies. In comparison, the signal frequency is increased, and thus closer to the minimum of the instrumental noise.

The entire duration of the MICROSCOPE mission is planned to be $18$ months and its launch is scheduled in 2016.

\subsection{The MICROSCOPE payload}
\label{ssec:2.2}

We have seen in the previous section that MICROSCOPE will be testing the Equivalence Principle by measuring the difference between the accelerations applied to maintain two masses of different composition on the same trajectory. This accelerations are measured thanks to an electrostatic accelerometer.

An electrostatic accelerometer is composed of a test mass surrounded by a set of electrodes. 
The association of the test mass with an electrode that faces it constitutes a capacitor. Any weak movement of the proof mass with respect to the electrode modifies the recovering surface or the gap between them and generates opposite variations of the relative capacitance. The difference of capacitance is detected through a charge amplifier and an heterodyne filtering, and the signal is digitized. The combination of the signals provided by the different electrodes - that correspond to different axes - provides the mass's position. 
The amplified signal is then  and processed with the control algorithm in order to compute the voltage to apply to the electrodes in order to compensate for the mass's motion and maintain it motionless at the center of the electrostatic cage. The computed voltage is amplified and opposite voltages are applied to a symmetric pair of electrodes in order to generate linear actuation forces. 
The fact that the same electrodes allow both the action and the detection of the mass' position is possible because of the difference of frequency bandwidths: the detection is performed with a $100 \, \mbox{kHz}$ pumping signal while the servo-loop channels exhibit frequency bandwidths of a few Hertz.
The generated voltage is proportional to the sensor acceleration \citep{Josselin99}. The proportionality factor depends on the voltage of the mass. In order to maintain this voltage to a constant value, the mass is connected to a $7 \, \mu \mbox{m}$ of diameter gold wire controlling its electrical potential. It is the only physical contact between the electrostatically levitated masses and the sensor cage.

\begin{figure}
  \includegraphics[width=0.5\textwidth]{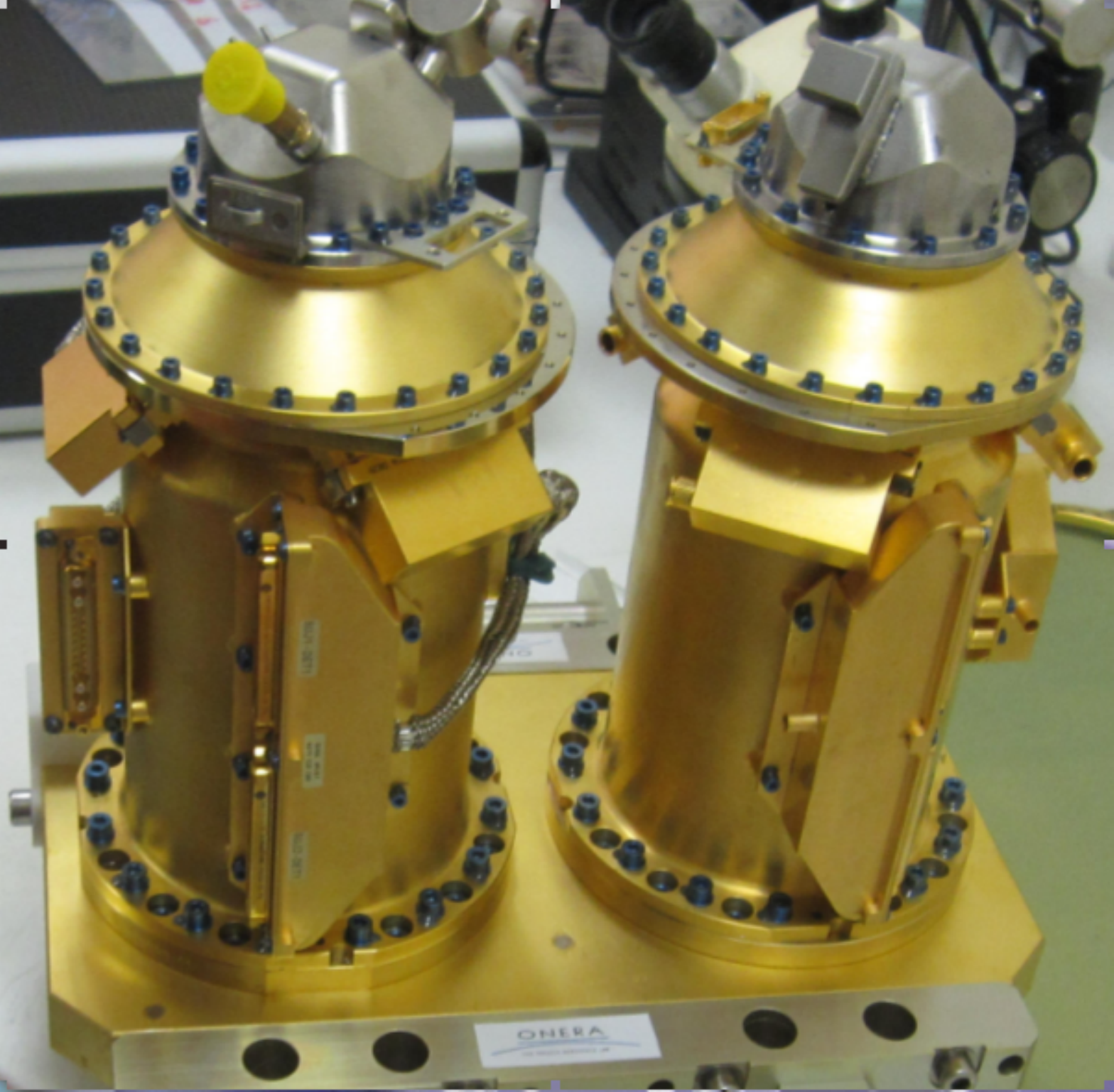}
  \includegraphics[width=0.5\textwidth]{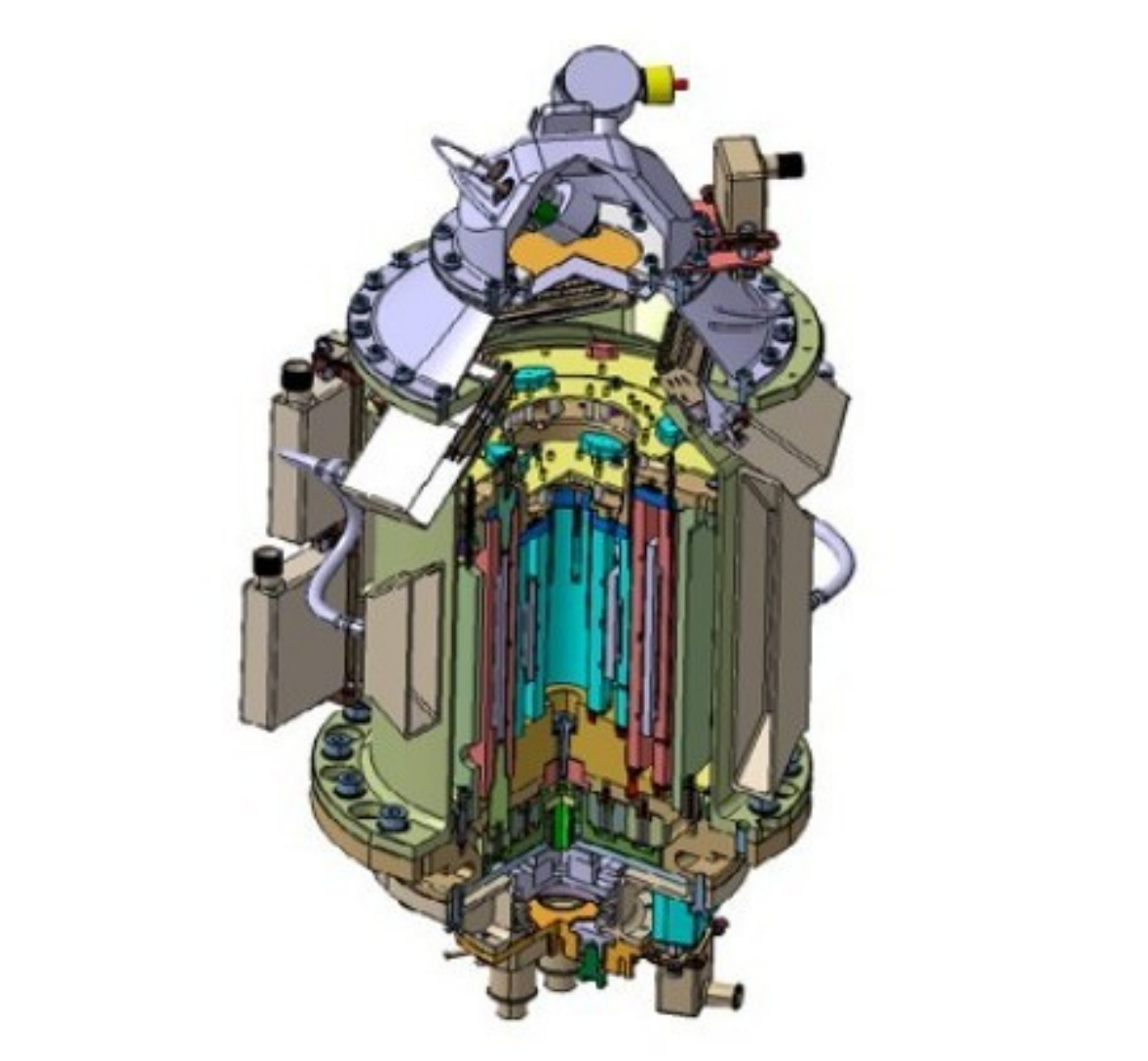}
\caption{{\itshape Left}: MICROSCOPE twin differential accelerometers in their tight housings, with the vacuum system on the top.
{\itshape Right}:  Each sensor unit is composed of two cylindrical concentric masses, here in purple, surrounded by electrodes, in blue.
}
\label{fig:schema-instrument}
\end{figure}

\begin{figure}
 \centering
  \includegraphics[width=1\textwidth,clip]{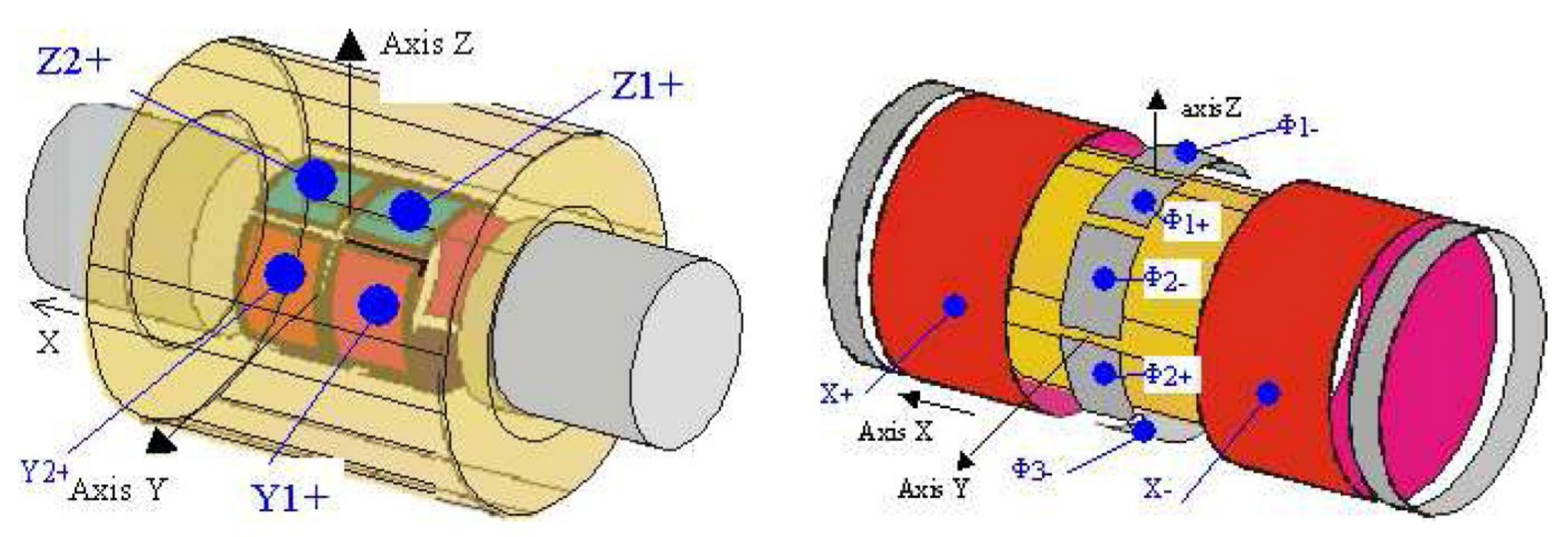}
\caption{The electrodes distribution around the mass: {\itshape Left}, the inner cylinder controls the radial axes; {\itshape Right}, the outer cylinder controls the sensitive axis.}
\label{fig:electrodes}
\end{figure}

The set of electrodes around each mass are engraved on two gold coated silica cylinders (see figure \ref{fig:electrodes}). Six pairs of electrodes enable the measurement of the mass' position and attitude and the control of its six degrees of freedom. The four electrodes of the inner cylinder control the radial axes $\vec{Y}$ and $\vec{Z}$ in translation and rotation. The outer cylinder is in charge of the ultra-sensitive $\vec{X}$ axis. The test of the Equivalence Principle is performed along this axis which is optimised to exhibit the best accuracy with a reduced electrostatic stiffness. The translation along this axis is controlled by the cylindrical outer electrodes positioned around the ends of the mass, while the rotation is controlled by the eight central quadrants of the outer cylinder in regard of four flat areas on the mass.

For the Equivalence Principle test, the accelerations of two masses of different composition are compared. Because the two masses are cylindrical and concentric, they have the same gravity center and are submitted to the same gravity field. The dimensions of the masses are chosen to provide the same moment of inertia along the three axes \citep{Lafargue-these}. The two masses and their electrodes constitute a differential accelerometer. The payload of the satellite, called T-SAGE (Twin Space Accelerometers for Gravitational Experimentation), consists in two independent and identical (except the composition of the masses) differential accelerometers. Developed by ONERA, these instruments benefit from the experience acquired during previous space missions such as GRACE and GOCE \citep{Touboul-SpaceAccelerometers}. The first instrument, which delivers the data to perform the test of the Equivalence Principle, includes one mass of Platinum Rhodium alloy (PtRh10) with 90\% Platinum and 10\% Rhodium, and one mass of Titanium alloy (TA6V) with 90\% Titanium, 6\% Aluminium and 4\% Vanadium. These materials have been selected among others (like technological reasons and macroscopic properties) because they have a large difference in subatomic particles, which may increase the intensity of the Equivalence Principle violation \citep{Damour-MaterialChoice}. The second differential accelerometer is composed of two masses constituted with the same material, PtRh10. The provided measurements are a reference to check the measurement exactitude.

Each differential electrostatic accelerometer is composed of three units: the Sensor Unit (SU), the Front End Electronic Unit (FEEU) and the Interface Control Unit (ICU). The SU corresponds to the mechanical core of the instrument: the test masses surrounded each by a set of electrodes arranged to perform the capacitive sensing of the mass motion and the control of the electrical fields generating the electrostatic actuation on the masses. The core is included in an Invar tight housing (see figure \ref{fig:schema-instrument}) required for the operation under vacuum better than $10^{-5} \, \mbox{Pa}$. The FEEU handles the analog electronics functions such as the capacitive sensors and the electrostatic actuations. It is linked to the ICU which hosts the digital electronics composed of 12 servo-loop channels, and delivers the instrument data to the satellite bus. The satellite payload operates in a finely stabilised thermal environment and is protected by a magnetic shield.

\section{The measurement principle}
\label{sec:3}

\subsection{The mass acceleration}
\label{ssec:3.1}

In order to extract a potential Equivalence Principle violation signal from the measurement, an accurate model of the acceleration measurement is required.

The proof mass $k$ is subject to the gravitational attraction of the Earth $\vec{g}$, to the electrostatic forces $\vec{fe_k}$ applied by the electrodes and to non-gravitational parasitic forces $\vec{F_{Pa,k}}$. Therefore, with the derivations performed in the satellite frame, the acceleration of the centre of mass $O_{k}$ of the mass $k$ reads as:
\begin{equation}
\ddot{\vec{OO_k}} = \frac{1}{m_{Ik}} \int_{P \in mass_k}{\vec{g}(P)dm_{gk}} + \frac{\vec{fe_k}}{m_{Ik}} + \frac{\vec{F_{Pa,k}}}{m_{Ik}} - [In] \cdot \vec{OO_k} - [Cor] \cdot \dot{\vec{OO_k}}
\end{equation}
with :
\begin{itemize}
  \item $O$ the origin of the selected inertial frame;
	\item $k$ the index of the mass (1 for the internal mass, 2 for the external mass);
	\item $m_{Ik}$ the inertial mass of the proof mass $k$;
	\item $m_{gk}$ the gravitational mass of the proof mass $k$;
	\item $[In]$ the inertia gradient matrix due to the satellite angular acceleration and velocity with respect to the inertial frame: 
	$$[In] \cdot \vec{OO_k} = \vec{\Omega} \wedge \vec{\Omega} \wedge \vec{OO_k} + \dot{\vec{\Omega}} \wedge \vec{OO_k} $$
	with $\vec{\Omega}$ the angular velocity of the satellite in the inertial frame;
	\item $[Cor]$ the matrix such that:
	$$ [Cor] \cdot \dot{\vec{OO_k}} = 2 \vec{\Omega} \wedge \dot{\vec{OO_k}} $$
\end{itemize}

The satellite is supposed to be a rigid body at the EP test frequency and is subject to the Earth gravity field, to the reaction to the electrostatic forces applied to each test mass and to the non-gravitational forces $\vec{F_{ng,sat}}$ which include the propulsion thrust $\vec{F_{th,sat}}$ and the external surface perturbations $\vec{F_{ext,sat}}$ such as the atmospheric drag and the solar and terrestrial radiation pressures. The acceleration of the center of mass $O_{sat}$ of the satellite is thus expressed as:
\begin{equation}
\ddot{\vec{OO_{sat}}} = \frac{1}{M_{Isat}} \int_{P \in sat}{\vec{g}(P)dM_{gsat}} + \frac{\vec{F_{ng,sat}}}{M_{Isat}} - [In] \cdot \vec{OO_{sat}} - [Cor] \cdot \dot{\vec{OO_{sat}}} - \sum_{j=1,2}{\frac{\vec{fe_j}}{M_{Isat}}}
\end{equation}
with:
\begin{itemize}
	\item $M_{Isat}$ the inertial mass of the satellite;
	\item $M_{gsat}$ the gravitational mass of the satellite;
\end{itemize}

So $\ddot{\vec{OO_{sat}}} - \ddot{\vec{OO_k}} = -\ddot{\vec{O_{sat}O_{k}}}$ and:
\begin{eqnarray}
\label{eq:acc-app}
\ddot{\vec{O_{k}O_{sat}}} & = & \frac{1}{M_{Isat}} \int_{P \in sat}{\vec{g}(P)dM_{gsat}} - \frac{1}{m_{Ik}} \int_{P \in mass_k}{\vec{g}(P)dm_{gk}}  \nonumber \\
                          &   & + R_{In,Cor}(\vec{O_{sat}O_{k}}) + \frac{\vec{F_{ng,sat}}}{M_{Isat}} - \frac{\vec{F_{Pa,k}}}{m_{Ik}} - \frac{\vec{fe_k}}{m_{Ik}} - \sum_{j=1,2}{\frac{\vec{fe_j}}{M_{Isat}}}
\end{eqnarray}
with
$$
R_{In,Cor}(\vec{O_{sat}O_{k}}) = [In] \cdot \vec{O_{sat}O_{k}} + [Cor] \cdot \dot{\vec{O_{sat}O_{k}}} + \ddot{\vec{O_{sat}O_{k}}}
$$

The Taylor series of the Earth gravity field at a point $P$ around the center of mass of the satellite $O_{sat}$ gives:
\begin{equation}
\vec{g}(P) = \vec{g}(O_{sat}) + [T] \cdot \vec{O_{sat}P} + \mathcal{O} \left( \vec{O_{sat}P} \right)
\end{equation}
where $[T]$ is the Earth gravity gradient tensor, whose components are on the order of $10^{-6} \, \mbox{s}^{-2}$ at the altitude of the experiment. The higher order derivatives are neglected, their values being lower than $10^{-12} \, \mbox{m}^{-1}\mbox{s}^{-2}$.
We have then: 
\begin{eqnarray}
\frac{1}{M_{Isat}} \int_{P \in sat}{\vec{g}(P)dM_{gsat}} - \frac{1}{m_{Ik}} \int_{P \in mass_k}{\vec{g}(P)dm_{gk}} \nonumber \\
\approx \left( \frac{M_{gsat}}{M_{Isat}} - \frac{m_{gk}}{m_{Ik}} \right) \vec{g}(O_{sat}) + \frac{m_{gk}}{m_{Ik}} [T]\vec{O_{sat}O_k} \nonumber \\
\end{eqnarray}
$\frac{m_{gk}}{m_{Ik}} - 1$ being lower than $10^{-12}$, we can approximate:
\begin{eqnarray}
\frac{1}{M_{Isat}} \int_{P \in sat}{\vec{g}(P)dM_{gsat}} - \frac{1}{m_{Ik}} \int_{P \in mass_k}{\vec{g}(P)dm_{gk}} \nonumber \\
\approx \left( \frac{M_{gsat}}{M_{Isat}} - \frac{m_{gk}}{m_{Ik}} \right) \vec{g}(O_{sat}) + [T]\vec{O_{sat}O_k} \nonumber \\
\end{eqnarray}

Let us define: $\vec{\Gamma_{App,k}} = \frac{\vec{fe_k}}{m_{Ik}}$. $\vec{\Gamma_{App,k}}$ is the measured acceleration deduced from the control voltages applied to the electrodes of the sensor core.

The applied acceleration on the mass is finally expressed by:
\begin{eqnarray}
\label{eq:applied}
\vec{\Gamma_{App,k}} &=& \left( \frac{M_{gsat}}{M_{Isat}} - \frac{m_{gk}}{m_{Ik}} \right) \vec{g}(O_{sat}) + ([T] - [In]) \cdot (\vec{O_{k}O_{sat}}) \nonumber \\
                     & & + \frac{\vec{F_{ng,sat}}}{M_{Isat}} - \frac{\vec{F_{Pa,k}}}{m_{Ik}} - \sum_{j=1,2}{\frac{\vec{fe_j}}{M_{Isat}}} \nonumber \\
                     & & + [Cor]\dot{\vec{O_{sat}O_{k}}} + \ddot{\vec{O_{sat}O_{k}}}
\end{eqnarray}
The relative motion of the proof mass with respect to the instrument frame linked to the satellite is servo-controlled to null to the accuracy of the loop, therefore the term $\ddot{\vec{O_{sat}O_{k}}}$ as well as the Coriolis term are negligible at the frequency of the Equivalence Principle test.

We define $\vec{\Gamma_{app,k}}$ so that:
\begin{equation}
\label{eq:gamma_app}
\vec{\Gamma_{App,k}} = \vec{\Gamma_{app,k}} + \frac{\vec{F_{ng,sat}}}{M_{Isat}} - \frac{\vec{F_{Pa,k}}}{m_{Ik}}
\end{equation}

\subsection{The inertial sensor measurement}
\label{ssec:3.2}

The physical parameters of the instrument perturb the measurement of $\vec{\Gamma_{App,k}}$; therefore the measured acceleration is not exactly equal to the applied acceleration:
\begin{eqnarray}
\label{eq:measurement}
\vec{\Gamma_{mes,k}} & = & \vec{b_{0,k}} + \left(\left[K_{1k}\right] + \left[\eta_k\right]\right) \cdot \left[\theta_k\right] \cdot \left(\vec{\Gamma_{app,k/sat}} + \frac{\vec{F_{ng,sat/sat}}}{M_{Isat}} \right) \nonumber \\
                     &   & + \left(\left[K_{1k} \right] + \left[\eta_k \right] \right) \cdot \left(-\frac{\vec{F_{Pa,k/instr}}}{m_{Ik}} -\frac{\vec{fe_{Pa,k/instr}}}{m_{Ik}} \right) \nonumber \\
                     &   & + K_{2k} \vec{\Gamma^2_{App,k/sat}} + \vec{\Gamma_{n,k}}
\end{eqnarray}
with
\begin{itemize}
  \item $\vec{b}_{0,k}$ the bias of the sensor $k$. It will be neglected in the following equations because the test of the Equivalence Principle is performed at the $f_{EP}$ frequency;
  \item $\vec{fe_{Pa,k}}$ represents the difference between the actually applied electrostatic force and its linear and quadratic considered model;
  \item $\left[ K_{2,k}\right]$ the quadratic diagonal matrix representing the major non linearities of the inertial sensor;
	\item $\vec{\Gamma^2_{App,k}}$, whose components are defined as the square values of the components of $\vec{\Gamma}_{App,k}$;
	\item $\vec{\Gamma}_{n,k}$ the measurement noise.
\end{itemize}

The sensitivity matrix $[M_k]$ represents by its antisymmetric part the misalignment parameters $\theta_{ij}$ ($\theta_{ij} = -\theta_{ji}$) which correspond to the small rotations between the reference frame of the satellite and the reference frame of the proof mass, i.e. the instrument. The symmetric part of $[M_k]$ represents the coupling between the axes $\eta_{ij}$ ($\eta_{ij} = \eta_{ji}$), for instance when the sensitive axes are not exactly orthogonal; the scale factors $K_{1ij}$ are the diagonal terms:
$$
[M_{k}] = \left[
\begin{array}{ccc}
K_{1xk}                 & \eta_{zk} + \theta_{zk} & \eta_{yk} - \theta_{yk} \\
\eta_{zk} - \theta_{zk} & K_{1yk}                 & \eta_{xk} + \theta_{xk} \\
\eta_{yk} + \theta_{yk} & \eta_{xk} - \theta_{xk} & K_{1zk}
\end{array}
\right]
$$
with $K_{1xk}-1 < 10^{-2}$, $\theta_{xk}$, $\theta_{yk}$ and $\theta_{zk} < 2.5 \times 10^{-3} \, \mbox{rad}$ and $\eta_{xk}$, $\eta_{yk}$ and $\eta_{zk} < 1 \times 10^{-4} \, \mbox{rad}$ constrained by the construction of the MICROSCOPE instrument and its accomodation on board the satellite.

We define common and differential contributions in the measured signal for both test-masses 1 and 2: 

\begin{equation}
\label{eq:common-diff}
\left\{
\begin{array}{c}
\vec{\Gamma_{mes,c}} = \frac{1}{2}(\vec{\Gamma_{mes,1}} + \vec{\Gamma_{mes,2}}) \\
 \\
\vec{\Gamma_{mes,d}} = \frac{1}{2}(\vec{\Gamma_{mes,1}} - \vec{\Gamma_{mes,2}})
\end{array}
\right.
\end{equation}

We define in the same way the common parameters, specified by the $c$ index, and the differential parameters, specified by the $d$ index.

The common measured acceleration $\vec{\Gamma_{mes,c}}$ is the input of the drag-free system, that forces the satellite thrusters to follow this measurement. So:
\begin{equation}
\vec{\Gamma_{mes,c}} = \vec{\Gamma_{res_{df}}} + \vec{C}
\end{equation}
where $\vec{C}$ is the drag-free command and $\vec{\Gamma_{res_{df}}}$ the drag-free loop residue.

The differential measured acceleration $\vec{\Gamma_{mes,d}}$ between the two masses contains the term $\delta = \frac{m_{g2}}{m_{I2}} - \frac{m_{g1}}{m_{I1}}$, which corresponds to the potential violation signal of the Equivalence Principle. Its expression is deduced from equation \ref{eq:measurement} and \ref{eq:common-diff}:
\begin{eqnarray}
\vec{\Gamma_{mes,d}} & = & \vec{b_{0d}} + \left(\left[K_{1c} \right] + \left[\eta_c \right] \right) \cdot \vec{b_{1d}} + \left(\left[K_{1c} \right] + \left[\eta_c \right] + \left[d\theta_c \right] \right) \cdot \vec{\Gamma_{app,d/sat}} \nonumber \\
                     &   & + \left(\left[K_{1d} \right] + \left[\eta_d \right] \right) \cdot \vec{b_{1c}} + \left(\left[K_{1d} \right] + \left[\eta_d \right] + \left[\theta_d \right] \right) \cdot \left(\vec{\Gamma_{App,c/sat}} - \vec{b_{1c}} \right) \nonumber \\
                     &   & + \frac{1}{2} K_{21} \vec{\Gamma^2_{App,1/sat}} - \frac{1}{2} K_{22} \vec{\Gamma^2_{App,2/sat}} + \frac{1}{2} \vec{\Gamma_{n,1}} - \frac{1}{2} \vec{\Gamma_{n,2}}
\end{eqnarray}
with
\begin{itemize}
  \item $\vec{b_{0d}}$ the difference of the read-out bias of the sensors
  \item $\vec{b_{1d}}$ the opposite of the half difference of the parasitic forces (including the electrostatic parasitic forces) applied on the two masses; $\vec{b_{1c}}$ is the half sum. Because of the slow variation of the parasitic forces, they can be considered as a bias at DC;
  \item $\left[d\theta_k \right]$ so that $\left[\theta_k \right] = I + \left[d\theta_k \right]$;
  \item $\vec{\Gamma_{n,1}}$, $\vec{\Gamma_{n,2}}$ the sensors read-out noise.
\end{itemize}

The expression of $\vec{\Gamma_{mes,c}}$ is in the same way:
\begin{eqnarray}
\label{eq:acc_c_mes}
\vec{\Gamma_{mes,c}} & = & \vec{b_{0c}} + \left(\left[K_{1d} \right] + \left[\eta_d \right] \right) \cdot \vec{b_{1d}} + \left(\left[K_{1d} \right] + \left[\eta_d \right] + \left[\theta_d \right] \right) \cdot \vec{\Gamma_{app,d/sat}} \nonumber \\
                     &   & + \left(\left[K_{1c} \right] + \left[\eta_c \right] \right) \cdot \vec{b_{1c}} + \left(\left[K_{1c} \right] + \left[\eta_c \right] + \left[d\theta_c \right] \right) \cdot \left(\vec{\Gamma_{App,c/sat}} - \vec{b_{1c}} \right) \nonumber \\
                     &   & + \frac{1}{2} K_{21} \vec{\Gamma^2_{App,1/sat}} + \frac{1}{2} K_{22} \vec{\Gamma^2_{App,2/sat}} + \frac{1}{2} \vec{\Gamma_{n,1}} + \frac{1}{2} \vec{\Gamma_{n,2}}
\end{eqnarray}

We deduce from equation \ref{eq:acc_c_mes} that $\vec{\Gamma_{App,c}}$, which is relatively weak because of this acceleration being nullified by the drag-free system, is equal to $[M_c]^{-1}(\vec{\Gamma_{mes,c}} - \vec{b_{0c}}) - [M_c]^{-1}[M_d]\vec{\Gamma_{app,d}}$ in first approximation.
The other terms are neglected: in the expression of $\vec{\Gamma_{mes,d}}$, $\vec{\Gamma_{App,c}}$ appears multiplied by a rather small matrix representing the disymmetries of the two sensors.

The expression of $\vec{\Gamma_{app,d}}$ is deduced from equation \ref{eq:applied} and \ref{eq:gamma_app}:
\begin{equation}
\vec{\Gamma_{app,d}} = \frac{1}{2} \delta \vec{g}(O_{sat}) + \frac{1}{2} ([T] - [In]) \cdot \vec{\Delta} 
\end{equation}
with $\vec{\Delta} = \vec{O_1O_2}$ the off-centring between the two masses.

Along the sensitive axis $\vec{X}$, the differential measured acceleration is therefore:

\begin{eqnarray}
\label{eq:Gamma-mesdx}
    \Gamma_{mes,dx} & \approx & \frac{1}{2}a_{c11} \, \delta \, g_x(O_{sat})                                     \nonumber \\
                    &         & + \frac{1}{2} {\left[
\begin{array}{c}
a_{c11} \\
a_{c12} \\
a_{c13}
\end{array}
\right]}^t [T - In] \left[
\begin{array}{c}
\Delta_x \\
\Delta_y \\
\Delta_z
\end{array}
\right]                                                                                                          \nonumber \\
                       &     & + {\left[
\begin{array}{c}
a_{d11} \\
a_{d12} \\
a_{d13}
\end{array}
\right]}^t (\vec{\Gamma}_{res_{df}} + \vec{C} - \vec{b_{0c}})                                                                   \nonumber \\
                       &    & +\: 2K_{2cxx} (\Gamma_{app,d} + b_{1dx}) \left(\frac{\Gamma_{res_{df,x}} + C_x - b_{0cx}}{K_{1cx}}\right)      \nonumber \\
                       &    & + K_{2dxx} \left((\Gamma_{app,d} + b_{1dx})^2 + \left(\frac{\Gamma_{res_{df,x}} + C_x - b_{0cx}}{K_{1cx}}\right)^2\right) \nonumber \\
                       &    & + b_{0dx} + \left([K_{1d}] + [\eta_d] \right) b_{1cx} + \left([K_{1c}] + [\eta_c] \right) b_{1dx} \nonumber \\
                       &    & + \Gamma_{n,1x} - \Gamma_{n,2x}
\end{eqnarray}

with
\begin{itemize}
	\item $a_{cij}$ the components of the matrix $\left[A_c\right] = [M_c] - [M_d] \cdot [M_c]^{-1} \cdot [M_d]$; at first order 
$\left[ \begin{array}{c}
a_{c11} \\
a_{c12} \\
a_{c13}\end{array} \right]^t \approx \left[ \begin{array}{c}
                                     K_{1cx} \\
                                     (\eta_{cz} + \theta{cz}) + \frac{K_{1dx}(\eta_{cy} - \theta_{cy})}{K_{1cx} K_{1cz}} \\
                                     (\eta_{cy} - \theta{cy}) + \frac{K_{1dx}(\eta_{cz} + \theta_{cz})}{K_{1cx} K_{1cy}}
                                     \end{array} \right]^t$;
	\item $a_{dij}$ the components of the matrix $\left[A_d\right] = [M_d] \cdot [M_c]^{-1}$; at first order 
$\left[ \begin{array}{c}
a_{d11} \\
a_{d12} \\
a_{d13}\end{array} \right]^t \approx \left[ \begin{array}{c}
                                     \frac{K_{1dx}}{K_{1cx}} \\
                                     \frac{(\eta_{dz} + \theta{dz})}{K_{1cy}} - \frac{K_{1dx}(\eta_{cz} + \theta_{cz})}{K_{1cx} K_{1cy}} \\
                                     \frac{(\eta_{dy} - \theta{dy})}{K_{1cz}} - \frac{K_{1dx}(\eta_{cy} - \theta_{cy})}{K_{1cx} K_{1cz}}
                                     \end{array} \right]^t$.
\end{itemize}

\section{In-orbit calibration}
\label{sec:4}

\subsection{Necessity of the in-orbit calibration}
\label{ssec:4.1}

The measurement accuracy is limited by different perturbation sources such as the effect of the gravity gradient, the attitude motion of the instrument, the residual acceleration of the orbital motion, the thermal and magnetic effects... At the considered altitude of $720 \, \mbox{km}$, the Earth gravity field is about $8 \, \mbox{ms}^{-2}$ and the $10^{-15}$ accuracy objective leads to allocate to all the contributors of the experience performance a level lower than $8 \times 10^{-15} \, \mbox{ms}^{-2}$.

In the equation of the differential measurement (\ref{eq:Gamma-mesdx}), several groups of errors explicitly arise, depending on the mass motion perturbations (parasitic forces or electrostatic disturbance), the AOCS servo-loops performances, the read-out electronics performances and the instrument characteristics, i.e. $\vec{\Delta}$, $a_{cij}$, $a_{dij}$ or $K_{2c}$ and $K_{2d}$. The contribution of this last group of error to the performance of the EP test has been evaluated \citep{Guiu-calib} taking into account the limitations of the manufacturing technology (see table \ref{tab:Budget}). It reaches $2 \times 10^{-13} \, \mbox{ms}^{-2}$ and is far too large compared to the EP test accuracy. This is why it is necessary to have an accurate calibration of those parameters, in order to correct the measurement of their effects. The calibration cannot be performed on-ground where the sensors are saturated, and must be performed in-orbit. For other missions, such as GReAT, it may be necessary to perform the calibration on-ground because of the the flight time duration, too short to enable an in-flight calibration \citep{Iafolla-GReAT}.

\begin{table}
\caption{Budget before calibration.}
\label{tab:Budget}
\begin{tabular}{lll}
\hline\noalign{\smallskip}
Disturbing term                                               & Concerned parameter                                        & Contribution to the            \\
                                                              & to be evaluated                                            & EP test measurement            \\
                                                              &                                                            & before calibration ($\mbox{ms}^{-2}$) \\
\noalign{\smallskip}\hline\noalign{\smallskip}
$a_{c11} \cdot T_{xx} \cdot \Delta_{x}$                       & $a_{c11} \cdot \Delta_{x} < 20.2 \, \mu \mbox{m} $         & $8.4 \times 10^{-14}$          \\
\cline{2-2}
$a_{c11} \cdot T_{xz} \cdot \Delta_{z}$                       & $a_{c11} \cdot \Delta_{z} < 20.2 \, \mu \mbox{m} $         & $8.6 \times 10^{-14}$          \\
\cline{2-2}
$a_{c11} \cdot T_{xy} \cdot \Delta_{y}$                       & $a_{c11} \cdot \Delta_{y} < 20.2 \, \mu \mbox{m} $         & $6 \times 10^{-16}$            \\
\cline{2-2}
$a_{c12} \cdot T_{yy} \cdot \Delta_{y} $                      & \begin{tabular}{l}
                                                                $a_{c12} < 2.6 \times 10^{-3} \, \mbox{rad}$ \\
                                                                $\Delta_{y} < 20 \, \mu \mbox{m}$
                                                                \end{tabular}
                                                                                                                           & $8.6 \times 10^{-16}$         \\
\cline{2-2}
$a_{c13} \cdot T_{zz} \cdot \Delta_{z} $                      & \begin{tabular}{l}
                                                                $a_{c13} < 2.6 \times 10^{-3} \, \mbox{rad}$ \\
                                                                $\Delta_{z} < 20 \, \mu \mbox{m}$
                                                                \end{tabular}
                                                                                                                           & $6.4 \times 10^{-16}$         \\
\cline{2-2}
$2 \cdot a_{d11} \cdot \Gamma_{res_{df},x}$                   & $a_{d11} < 10^{-2} $                                       & $2 \times 10^{-14}$           \\
\cline{2-2}
$2 \cdot a_{d12} \cdot \Gamma_{res_{df},y}$                   & $a_{d12} < 1.6 \times 10^{-3} \, \mbox{rad}$               & $3.0 \times 10^{-15}$         \\
\cline{2-2}
$2 \cdot a_{d13} \cdot \Gamma_{res_{df},z}$                   & $a_{d13} < 1.6 \times 10^{-3} \, \mbox{rad}$               & $3.0 \times 10^{-15}$         \\
\cline{2-2}
$4 \cdot K_{2,cxx} \cdot \Gamma_{app,dx} \cdot \Gamma_{res_{df},x}$ & $K_{2,cxx} < 14000 \, \mbox{s}^2/\mbox{m}$           & $8.0 \times 10^{-16}$         \\
\cline{2-2}
$2 \cdot K_{2,dxx} \cdot (\Gamma_{app,dx}^2 + \Gamma_{res_{df},x}^2)$ & $K_{2,dxx} < 14000 \, \mbox{s}^2/\mbox{m}$         & $8.0 \times 10^{-16}$         \\
\noalign{\smallskip}\hline\noalign{\smallskip}
Total                                                         &                                                            & $2 \times 10^{-13}$           \\
\noalign{\smallskip}\hline
\end{tabular}
\end{table}

Other perturbations, such as the effect of the gravity gradient or the thermal and magnetic effects, are taken into account in the global error budget \citep{Touboul-2009}. In order to reach the accuracy objective for the EP test, it has been determined that the residue of error due to the instrumental parameters after the complete in-orbit calibration and the measurement correction should be limited to less than $2 \times 10^{-15} \, \mbox{ms}^{-2}$. For this purpose ten parameters have been identified to be evaluated: $a_{c11} \Delta_x$, $a_{c11} \Delta_y$ and $a_{c11} \Delta_z$; $a_{c12}$ and $a_{c13}$; $a_{d11}$, $a_{d12}$ and $a_{d13}$; $K_{2cxx}$ and $K_{2dxx}$. The required calibration accuracy of these parameters is presented in table \ref{tab:perfo-numerical}.

\begin{table}
\scriptsize
\caption{Numerical performance of the calibration procedures: results on a set of 100 simulations.}
\label{tab:perfo-numerical}
\begin{tabular}{lcccc}
\hline\noalign{\smallskip}
Parameter to be            & Objectives on the                & Worst estimation                 & Mean estimation                  & Standard deviation of the          \\
calibrated                 & accuracy of the                  & accuracy after                   & accuracy after                   & estimation accuracy                \\
                           & estimation                       & calibration                      & calibration                      & after calibration                  \\
\noalign{\smallskip}\hline\noalign{\smallskip}
$a_{c11} \cdot \Delta_{x}$ & $0.1 \, \mu \mbox{m}$            & $0.04 \, \mu \mbox{m}$           & $0.01 \, \mu \mbox{m}$           & $7.3 \, \mbox{nm}$                 \\
$a_{c11} \cdot \Delta_{z}$ & $0.1 \, \mu \mbox{m}$            & $0.05 \, \mu \mbox{m}$           & $0.03 \, \mu \mbox{m}$           & $6.5 \, \mbox{nm}$                 \\
$a_{c11} \cdot \Delta_{y}$ & $2 \, \mu \mbox{m}$              & $0.2 \, \mu \mbox{m}$            & $0.05 \, \mu \mbox{m}$           & $0.04 \, \mu \mbox{m}$             \\
$a_{c12}$                  & $9.0 \times 10^{-4}\,\mbox{rad}$ & $1.0 \times 10^{-3}\,\mbox{rad}$ & $2.6 \times 10^{-4}\,\mbox{rad}$ & $2.3 \times 10^{-4} \, \mbox{rad}$ \\
$a_{c13}$                  & $9.0 \times 10^{-4}\,\mbox{rad}$ & $1.1 \times 10^{-3}\,\mbox{rad}$ & $3.1 \times 10^{-4}\,\mbox{rad}$ & $2.6 \times 10^{-4} \, \mbox{rad}$ \\
$a_{d11}'$                 & $1.5 \times 10^{-4}$             & $5 \times 10^{-6}$               & $1.6 \times 10^{-6}$             & $1.2 \times 10^{-6}$               \\
$a_{d12}$                  & $5 \times 10^{-5} \, \mbox{rad}$ & $2 \times 10^{-5} \, \mbox{rad}$ & $1.2 \times 10^{-6}\,\mbox{rad}$ & $1.7 \times 10^{-6} \, \mbox{rad}$ \\
$a_{d13}$                  & $5 \times 10^{-5} \, \mbox{rad}$ & $4 \times 10^{-6} \, \mbox{rad}$ & $1.0 \times 10^{-6}\,\mbox{rad}$ & $8.3 \times 10^{-7} \, \mbox{rad}$ \\
$K_{2dxx}/K_{1cx}^2$       & $250 \, \mbox{s}^2/\mbox{m}$     & $124 \, \mbox{s}^2/\mbox{m}$     & $25 \, \mbox{s}^2/\mbox{m}$      & $23 \, \mbox{s}^2/\mbox{m}$        \\
$K_{2cxx}/K_{1cx}^2$       & $1000 \, \mbox{s}^2/\mbox{m}$    & $274 \, \mbox{s}^2/\mbox{m}$     & $62 \, \mbox{s}^2/\mbox{m}$      & $54 \, \mbox{s}^2/\mbox{m}$        \\
\noalign{\smallskip}\hline
\end{tabular}
\end{table}

\subsection{The calibration process}
\label{ssec:4.2}

The central idea of the calibration is to create an acceleration signal that amplifies the effect of the parameter to be determined, the corresponding term becoming predominant in the measurement equation. The calibration signal is obtained by using on one hand the capability of the Attitude and Orbit Control System (AOCS) to force the motion of the satellite through its thrusters and on the other hand the capability of the electrostatic control loop to force the motion of the test masses \citep{Guiu-these}. The measurement acceleration, which is the input of the drag-free system, is available; therefore the amplitude of the signal does not require to be accurately known, contrary to its frequency and its phase.

The signal to be detected is composed of a systematic and a stochastic contribution. This second contribution is induced by the accelerometer noise and the stochastic variations of the measurement components. To reduce this stochastic error to an acceptable level, the calibration duration of each parameter is fixed to 10 orbits. 

The ten parameters to be calibrated are gathered in table \ref{tab:perfo-numerical}. A specific method is proposed for each parameter in order to amplify its effect.

\begin{description}

	\item[{\bfseries Off-centring of the proof masses along the $\vec{X}$ and $\vec{Z}$ axes:}] The off-centring $\Delta_{x}$ between the two test masses introduces an effect coming from the Earth gravity gradient and the inertia gradient matrix (see equation \ref{eq:Gamma-mesdx}). The parameter $a_{c11} \Delta_{x}$ appears in the equation of the differential measurement along the ultra-sensitive axis $\vec{X}$ (equation \ref{eq:Gamma-mesdx}) in the term $a_{c11} \cdot T_{xx} \cdot \Delta_{x}$. The Earth gravity gradient $T_{xx}$ is a very well-known signal of strong amplitude at two times the orbital frequency $f_{orb}$ when the satellite is in inertial pointing. At $2f_{orb}$, this term is predominant in the measurement equation \citep{Touboul-2012}, and the calibration equation is:
\begin{equation}
\Gamma_{mes,dx/cos}(2f_{orb}) \approx \Gamma_{calib_1} = \frac{1}{2} T_{xx}(2f_{orb}) \cdot a_{c11} \cdot \Delta_{x}
\end{equation}
The term containing the parameter to be calibrated is extracted; all the other terms at $2f_{orb}$ participate to the error on the parameter determination.
	
With the same method, the sine phase part of the measurement signal is used to estimate $a_{c11} \Delta_{z}$:
\begin{equation}
\Gamma_{mes,dx/sin}(2f_{orb}) \approx \frac{1}{2} T_{xz}(2f_{orb}) \cdot a_{c11} \cdot \Delta_{z}
\end{equation}
\\

\item[{\bfseries Off-centring of the proof masses along the $\vec{Y}$ axis:}] The previous method used to estimate the off-centrings along the $\vec{X}$ and $\vec{Z}$ axes cannot be used for $a_{c11} \Delta_{y}$ because of the weak value of the corresponding Earth gravity gradient component $ T_{xy}$.
To amplify the effect of this off-centring, the AOCS can force the motion of the satellite around the $\vec{Z}$ axis at $f_{cal/ang}$ with an amplitude $\alpha_0$ and therefore creates an additional angular acceleration introduced in the measurement equation through the inertia gradient matrix. This oscillation also induces a non null value of $T_{xy}$ at $f_{cal/ang}$.
\begin{equation}
\Gamma_{mes,dx}(f_{cal/ang}) \approx \frac{1}{2} \left(T_{xy}(f_{cal/ang}) - \alpha_0 \cdot \omega_{cal/ang}^2 \right) \cdot a_{c11} \cdot \Delta_{y}
\end{equation}
\\

\item[{\bfseries Parameters of the common sensitivity matrix:}] The misalignment and coupling terms $a_{c12}$ and $a_{c13}$ multiply the Earth gravity gradient. It is therefore possible to use the same signal as for $a_{c11} \Delta_{x}$ and $a_{c11} \Delta_{z}$, the naturally strong Earth gravity gradient at $2f_{EP}$ when the satellite is in inertial pointing, associated with an oscillation of the masses, in order to discriminate between the estimation of the off-centring parameters and the estimation of the misalignment parameters. The test masses therefore oscillate along the $\vec{Y}$ axis for $a_{c12}$  and along the $\vec{Z}$ axis for $a_{c13}$ at the frequency $f_{TM}$ and with the amplitude $\Delta_{TM}$. The oscillating mass motion modulates the Earth gravity gradient signal, leading to measurable effects at $f_{TM} \pm 2f_{orb}$.

The calibration signal for $a_{c12}$ is:
\begin{equation}
\Gamma_{mes,dx/cos}(f_{TM} \pm 2f_{orb}) \approx \frac{1}{2} \left(- a_{c11} \cdot T_{xx}(2f_{orb}) \right) \cdot \Delta_{TM} \cdot a_{c12}
\end{equation}
The calibration equation for $a_{c13}$ is:
\begin{equation}
\Gamma_{mes,dx/cos}(f_{TM} \pm 2f_{orb}) \approx \frac{1}{2} \left(T_{zz}(2f_{orb}) - a_{c11} \cdot T_{xx}(2f_{orb}) \right) \cdot \Delta_{TM} \cdot a_{c13}
\end{equation}

Since $T_{zz}(2f_{orb}) \approx -T_{xx}(2f_{orb})$, the signal to be extracted for the calibration of $a_{c12}$ is half as strong as the signal for the calibration of $a_{c13}$. The noise being the predominant error term and decreasing as the square root of the integration time, the calibration session must last four times longer to obtain the same accuracy. The calibration session for the estimation of $a_{c12}$ will thus last for 40 orbits instead of 10. 
\\

\item[{\bfseries Parameters of the differential sensitivity matrix:}] 
In equation \ref{eq:Gamma-mesdx}, the differential parameters are in factor of the common mode acceleration, which is the input of the drag-free system. The effect of $a_{d11}$, $a_{d12}$ and $a_{d13}$ are therefore amplified by adding a secondary input to the drag-free control: a command $C_x$ at a well-known frequency. The satellite therefore oscillates along the $\vec{X}$ axis for $a_{d11}$, the $\vec{Y}$ axis for $a_{d12}$ and the $\vec{Z}$ axis for $a_{d13}$, following a sine motion at the frequency $f_{cal/lin}$.

In fact, $a_{d11}'$ is estimated instead of $a_{d11}$, with:
\begin{equation}
a_{d11}' = a_{d11} + 2\frac{K_{2cxx}}{K_{1cx}}b_{1dx} + 2\frac{K_{2dxx}}{K_{1cx}}(C_{0x} - b_{0cx})
\end{equation}
The purpose is the gathering in this parameter of all the biases in factor of $C_x$ in equation \ref{eq:Gamma-mesdx} (except for the component of $\vec{\Gamma_{app,d}}$ at DC, which is supposed to be negligible over the ten orbits), and which therefore contribute to the measured acceleration at $f_{cal/lin}$.

The calibration equation for $a_{d11}'$ is:
\begin{equation}
\Gamma_{mes,dx}(f_{cal/lin}) \approx \Gamma_{mes,cx}(f_{cal/lin}) \cdot a_{d11}'
\end{equation}
It is similar for $a_{d12}$:
\begin{equation}
\Gamma_{mes,dx}(f_{cal/lin}) \approx \Gamma_{mes,cy}(f_{cal/lin}) \cdot a_{d12}
\end{equation}
And $a_{d13}$:
\begin{equation}
\Gamma_{mes,dx}(f_{cal/lin}) \approx \Gamma_{mes,cz}(f_{cal/lin}) \cdot a_{d13}
\end{equation}
\\

\item[{\bfseries Differential quadratic factor:}] As for $a_{d11}'$, the drag-free system commands a sine motion of the satellite along the $\vec{X}$ axis at $f_{cal/lin}$. The detected acceleration is extracted at $2f_{cal/lin}$ and is proportional to $\frac{K_{2dxx}}{K_{1cx}^2}$:
\begin{equation}
\Gamma_{mes,dx}(2f_{cal/lin}) \approx \Gamma_{mes,cx}^2(2f_{cal/lin}) \cdot \frac{K_{2dxx}}{K_{1cx}^2}
\end{equation}
\\

\item[{\bfseries Common quadratic factor:}] $\frac{K_{2cxx}}{K_{1cx}^2} = \frac{\frac{1}{2}(K_{21xx} + K_{22xx})}{K_{1cx}^2}$ is obtained from the calibration of $K_{21xx}$ and $K_{22xx}$. The calibration is done in two steps. First, the quadratic factor $K_{21xx}$ of the sensor $1$ is estimated by forcing a sine motion of the test mass $1$ along the $\vec{X}$ axis at a frequency $f_{TM}'$ with an amplitude $\Delta_{TM}$, while the satellite drag-free system is locked on the test mass $2$:
\begin{equation}
\Gamma_{mes,1x}(2f_{TM}') \approx \Gamma_{mes,1x}^2(2f_{TM}') \cdot \frac{K_{21xx}}{K_{11x}^2}
\end{equation}
Then $K_{22xx}$ is estimated in the same way but the role of the two test masses is permuted. Two sessions of ten orbits are thus necessary.
When both quadratic factors have been calibrated, the common parameter $K_{2cxx}$ is inferred.
\\
\end{description}

The performances of these calibration procedures have been first evaluated analytically by developing the expression of the measurement at the first order \citep{Levy-calibration}. They are compatible with the requirements of the MICROSCOPE mission. Nevertheless, the system to be analyzed includes several servo-loops and the implementation of a software simulator is necessary in order to validate them numerically.

\section{The calibration simulator}
\label{sec:5}

\subsection{Design of the calibration simulator}
\label{ssec:5.1}

The purpose of this software is to simulate the measurement equation during the calibration sessions, as well as the data processing used to extract the parameters' estimations.

\begin{figure}
 \centering
  \includegraphics[width=1\textwidth,clip]{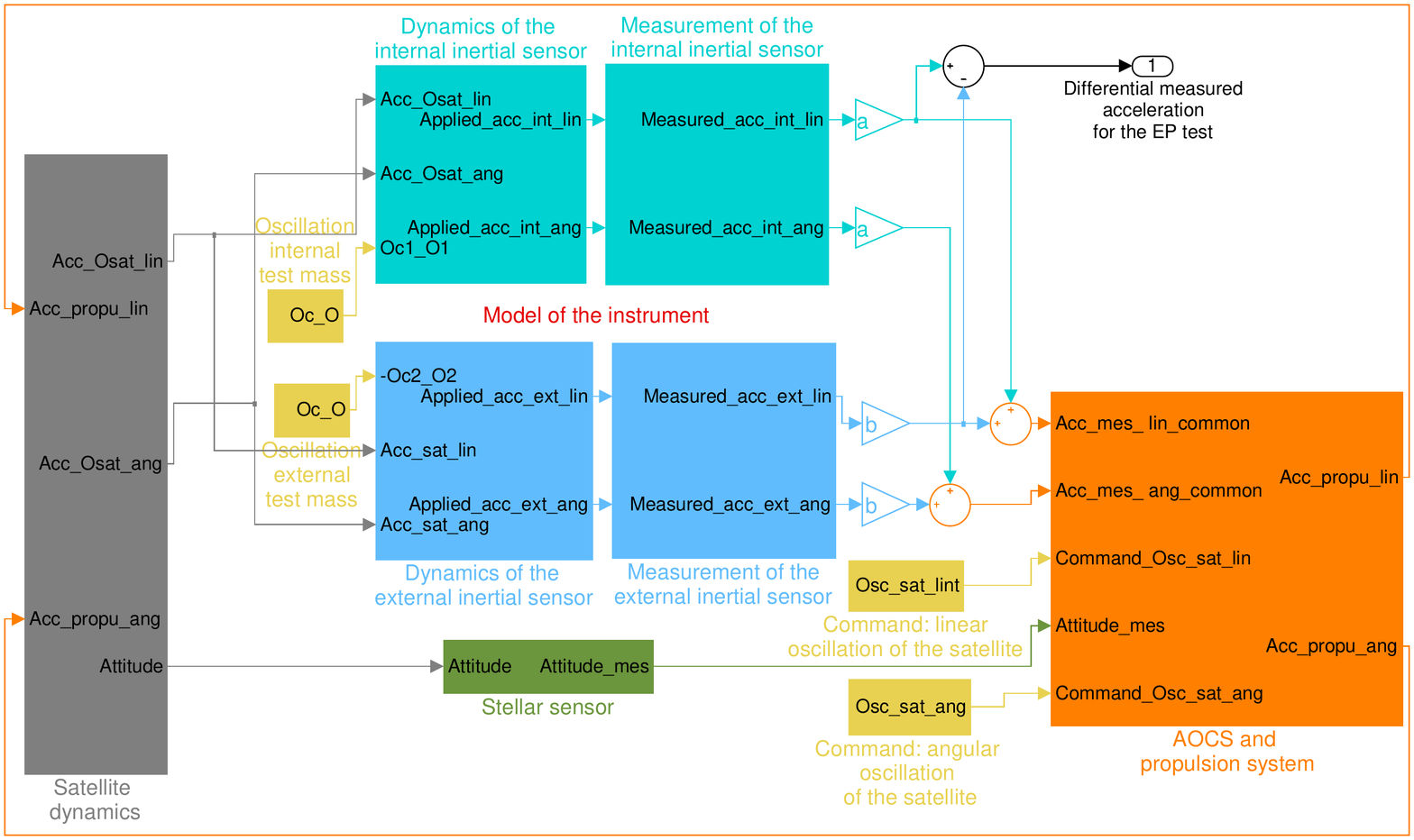}
\caption{Scheme of the calibration simulator. The secondary inputs used for the calibration procedures are in yellow.}
\label{fig:simulator}
\end{figure}

The calibration simulator has been developed with Simulink. It recreates the dynamics of the satellite in its environment, the drag-free loop and the instrument. The instrumental parameters are set to initial realistic values. A random process allows the selection of a set of parameters, and statistics results can be deduced from numerous simulations. The calibration sessions correspond to the conditions described in the procedures seen in the previous part: linear or angular oscillation of the satellite, oscillation of the test masses or observation of the measured Earth gravity gradient. The calibration signal is initialized in the simulator with its specified amplitude and frequency depending on the parameter to be calibrated, and is injected as secondary input in the accelerometer loop or in the AOCS. The simulator is composed of several blocks modelling several subsystems (see figure \ref{fig:simulator}).

\subsubsection{Satellite dynamics in a perturbing environment}

The `satellite dynamics' block provides non-gravitational acceleration of the satellite. It accounts for the thrust applied by the propulsion system, to which are added the non-gravitational external perturbations. They are introduced in the simulator from a data file that represents the non-gravitational atmospheric residual drag and the solar radiation pressure accelerations for the expected specific trajectory and orientation of the satellite.

\subsubsection{The instrument}

The instrument corresponds to only one of the two differential electrostatic accelerometers with its inner and outer test masses. The control loop of the accelerometer is not represented here, considering that the cut-off frequency are sufficiently large with respect to the calibration frequencies ($10^{-2} \, \mbox{Hz}$): this loop includes an integral term in the control law to suppress the dynamics error at the lower frequencies. The model of each sensor is composed of two parts: the first one computes the electrostatic acceleration applied to the inertial sensor mass to counteract its motion relatively to the satellite; the second one simulates the measurement of this applied acceleration.

The `dynamics' block computes the acceleration $\vec{\Gamma_{App,k}}$ applied to the mass $k$ as deduced from equation \ref{eq:applied}:
\begin{equation}
\label{eq:instrument-dynamics}
\vec{\Gamma_{App,k}} = \vec{\Gamma}(O_{sat}) + ([T] - [In]) \cdot \vec{O_{k}O_{sat}} - \frac{F_{Pa,k}}{m_{Ik}} - [Cor] \cdot \dot{\vec{O_{k}O_{sat}}} - \ddot{\vec{O_{k}O_{sat}}}
\end{equation}
with
\begin{itemize}
	\item $\vec{\Gamma(O_{sat})}$ the acceleration of the satellite. It only represents the non-gravitationnal acceleration, because the potential violation of the Equivalence Principle is neglected for the calibration simulator, and the orbital motion term therefore disappears in the expression of $\vec{\Gamma_{App,k}}$ in equation \ref{eq:applied};
	\item $O_{k}$ the centre of mass of the test mass $k$.
\end{itemize}

The off-centring between the satellite's center of mass and the test mass's center of mass induces effects due to the Earth gravity gradient $[T]$ and the inertia gradient matrix $[In]$ if the satellite rotates. Moreover, an additional Coriolis effect appears if the test mass is in motion relatively to the rotating satellite.

\begin{figure}
 \centering
  \includegraphics[width=0.8\textwidth,clip]{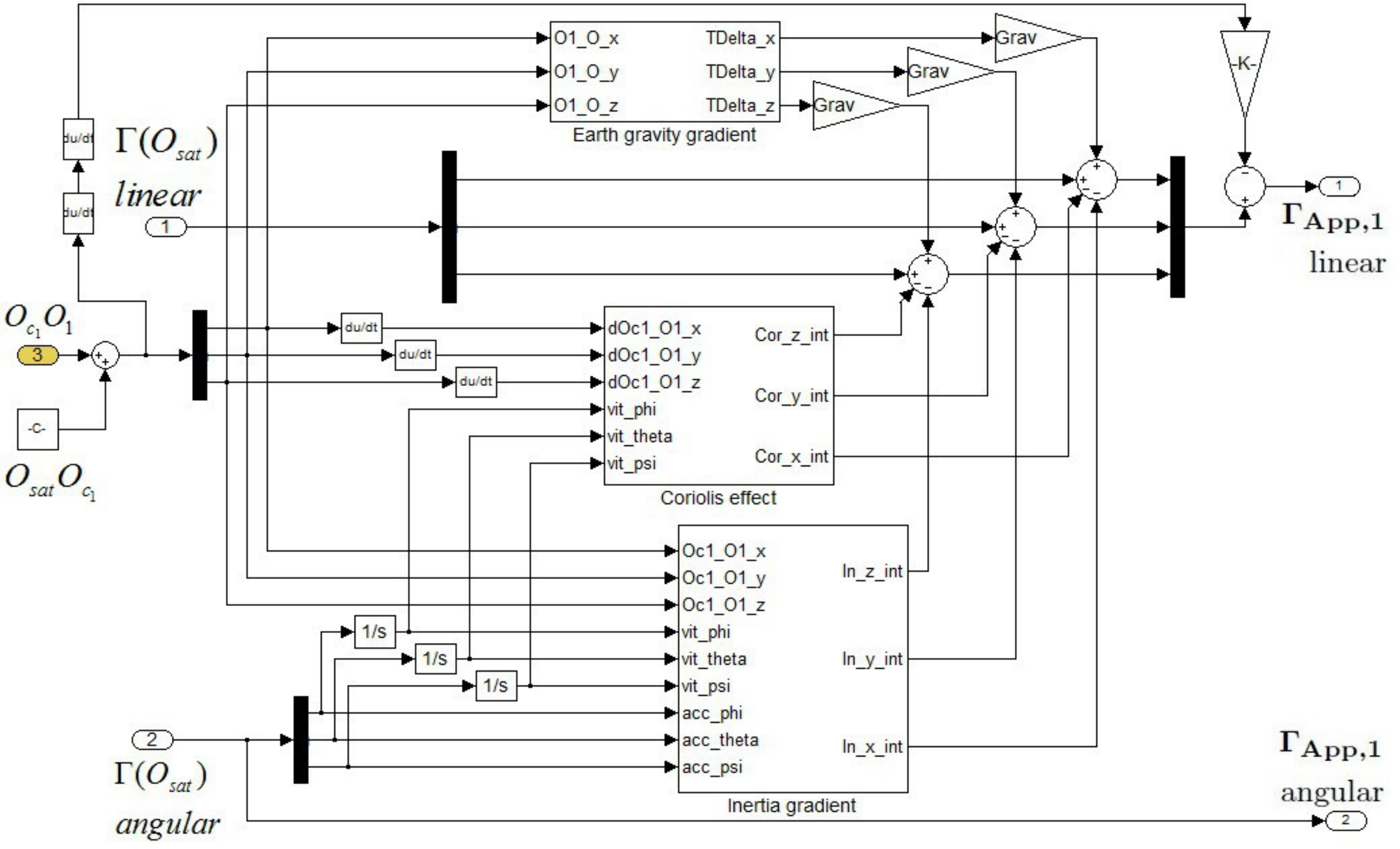}
\caption{Model of the internal sensor ($1$) dynamics}
\label{fig:instrument-dynamics}
\end{figure}

Equation \ref{eq:instrument-dynamics} is implemented in the simulator, as shown in figure \ref{fig:instrument-dynamics}. The six axes of the accelerometer are represented.

\begin{figure}
 \centering
  \includegraphics[width=0.8\textwidth,clip]{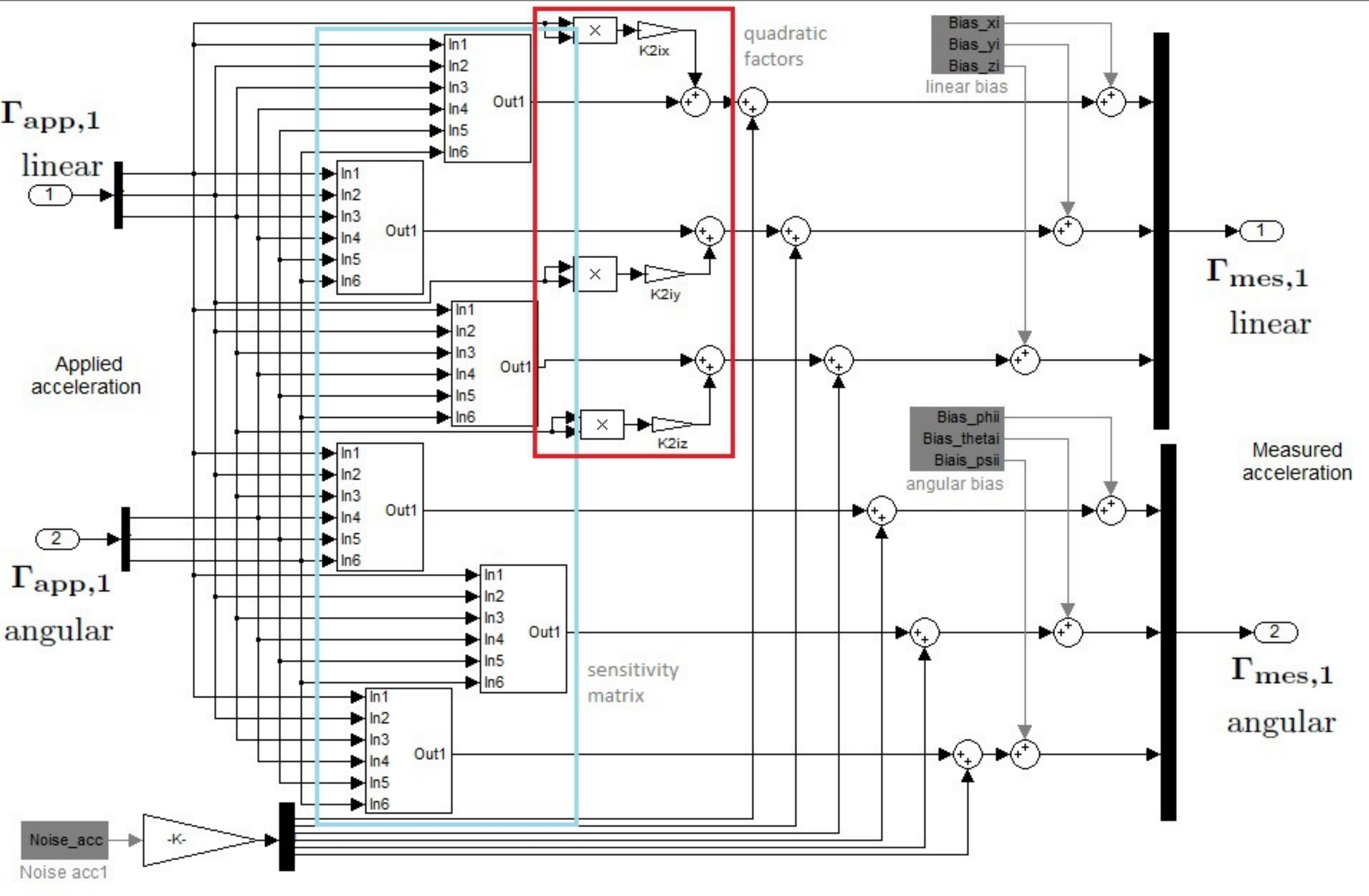}
\caption{Model of the internal sensor ($1$) measurement}
\label{fig:instrument-measurement}
\end{figure}

In the case of a perfect instrument, the measured acceleration would be equal to the applied acceleration. However, some instrumental parameters limit the measurement accuracy: noise, bias, sensitivity matrix and quadratic non linearities. They are represented in the second block of the sensor model according to equation \ref{eq:measurement} as shown in figure \ref{fig:instrument-measurement}. 

The output of the measurement block is the measured acceleration $\Gamma_{mes,k}$, which leads to the common and the differential measured accelerations. The first one becomes the input of the AOCS block while the second one is processed for the EP test experiment.

\subsubsection{AOCS and propulsion system}

The AOCS computes the thrust to be applied to the satellite to compensate for the surface perturbations and thereby minimize the non-gravitational acceleration measured by the inertial sensor. In addition, it also compensates for the bias $b_{0c}$ of the accelerometer. The propulsion command $\vec{\Gamma_{DF}}$ is: 

\begin{equation}
\vec{\Gamma_{DF}} = TF_{DF} \cdot (\vec{\Gamma_{mes,c}} + \vec{C_{cal}})
\end{equation}
with
\begin{itemize}
	\item $\vec{C_{cal}}$ the calibration signal for the linear oscillation of the satellite;
	\item $TF_{DF}$ the transfer function of the AOCS.
\end{itemize}

The additional calibration signal is used for the estimation of the differential parameters $a_{d11}'$, $a_{d12}$, $a_{d13}$ and $K_{2dxx}$ - the satellite oscillates along the different axes.

Two different transfer functions are used for the linear and the angular accelerations; both are provided by the CNES team in charge of this satellite sub-system.

A star tracker provides the AOCS with the angular position of the satellite. In fact two optical sensors are used with orthogonal field of view axes. The model of this equipment in the simulator takes its noise and bias into account.

For the linear satellite acceleration, the differential accelerometer is the only sensor. On the opposite, as shown in figure \ref{fig:simulator}, the AOCS angular loop is composed of two different chains: one running through the instrument and the other through the star tracker. The star tracker measures the satellite attitude while the instrument processes its angular acceleration. The estimate of the satellite attitude which feeds the AOCS is obtained by combining the double integration of the angular acceleration measurement with the measurement of the satellite attitude. This hybridization process improves the attitude accuracy because the instrument delivers a precise measurement at high frequencies, while the star tracker is better about DC. Two complementary filters are used to estimate the angular measurement: a high-pass filter applied to the accelerometer measurement in order to eliminate the bias, and a low-pass filter applied to the star tracker measurement.

The calibration signal for the angular oscillation of the satellite is added as a secondary input on the star tracker attitude measurement. It allows the calibration of $a_{c11} \Delta_{y}$.

Once the compensating acceleration $\vec{\Gamma_{DF}}$ is calculated, it is applied to the satellite by the propulsion system. This system is composed of several cold gas micro-thrusters using nitrogen, set at the four corners of the cubic external structure of the satellite. The performance of the propulsion is limited by imperfections represented in the simulator with a noise $\vec{\Gamma_{n,DF}}$ and a sensitivity matrix $[M_{DF}]$, corresponding to the imperfect knowledge of the thrust axes. The corrective acceleration applied by the thrusters is therefore not exactly equal to the command, but is:
\begin{equation}
\vec{\Gamma_{propu}} = -[M_{DF}] \cdot \vec{\Gamma_{DF}} + \vec{\Gamma_{n,DF}}
\end{equation}

\begin{figure}
 \centering
  \includegraphics[width=1\textwidth,clip]{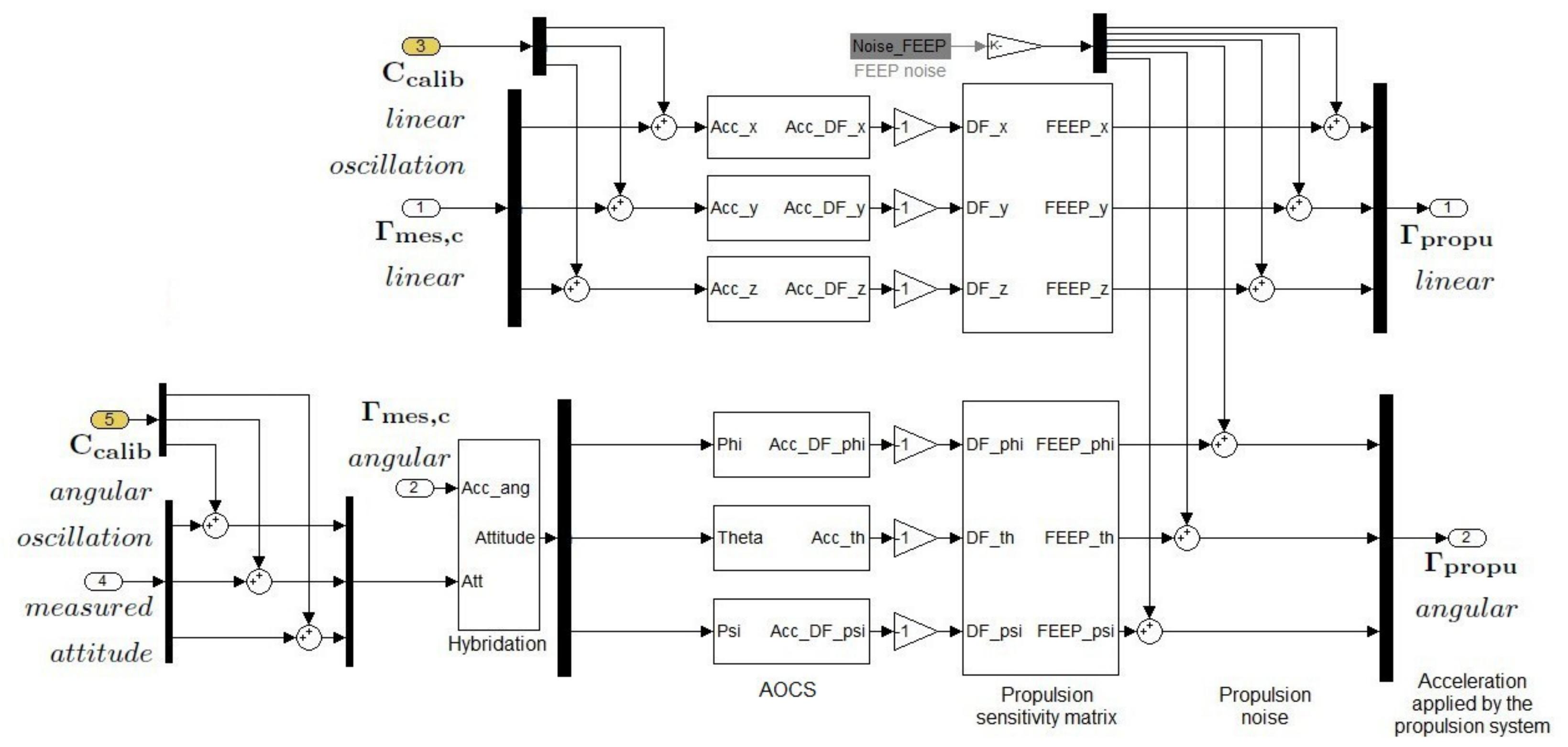}
\caption{Model of the AOCS and propulsion system}
\label{fig:SCAA-propulsion}
\end{figure}

The implemented model of the AOCS and the propulsion system is presented in figure \ref{fig:SCAA-propulsion}.

\subsection{Exploitation of the simulator}
\label{ssec:5.2}

\subsubsection{Data processing}

The calibration simulator generates acceleration measurements every $0.25 \, \mbox{s}$ (in-orbit sampling rate of the instrument) for the different calibration sessions of ten orbits. The measurement data are then processed to estimate the parameter values as it will be done during the mission. 

A pass band filter around the calibration frequency composed of two second order Butterworth filters is first applied to the data in the time domain. The Least Square method is then used to extract the signal at the calibration frequency after eliminating the 200 first seconds corresponding to the transient phase of the filter.

Each parameter is evaluated separately after each calibration sequence. After a first round of estimation, the estimation errors can be reduced by reprocessing the data: the exploited measurements used for the calibration are precisely corrected with the already estimated value of the parameters. Each iteration improves the global accuracy, as the estimation of each individual parameter benefits from the refinement of the others. For example, to estimate the off-centring along $\vec{X}$, $a_{c11} \Delta_{x}$, we use the relation $\vec{\Gamma_{res_{df}}} = \vec{\Gamma_{mes,c}}$ at all non-zero frequencies to define the new calibration equation:
\begin{eqnarray}
\Gamma_{calib_2} & = & 2\Gamma_{mes,dx/cos}(2f_{orb}) \nonumber \\
                 &   & - \hat{a_{c13}} \cdot \hat{T_{zz}}(2f_{orb}) \cdot \hat{\Delta_{z}} \nonumber \\
                 &   & - 2 \hat{a_{d11}'} \cdot \Gamma_{mes,cx}(2f_{orb}) \nonumber \\
                 &   & - 2 \hat{a_{d12}} \cdot \Gamma_{mes,cy}(2f_{orb}) \nonumber \\
                 &   & - 2 \hat{a_{d13}} \cdot \Gamma_{mes,cz}(2f_{orb}) \nonumber \\
                 & = & \hat{T_{xx}}(2f_{orb}) \cdot a_{c11} \cdot \Delta_{x}
\end{eqnarray}
where the hat symbol is associated to an estimated value.

\subsubsection{Results}

A set of 100 simulations has been run, each test corresponding to a random draw of the parameters to be calibrated. The ten parameters have been calibrated for each test, corresponding to a 140 orbits time span.

The worst estimation accuracy obtained for each parameter as well as the mean results and the standard deviation are gathered in table \ref{tab:perfo-numerical}. The results are compatible with the objectives, except for the common misalignment parameters with respect to $\vec{Y}$ and $\vec{Z}$, $a_{c13}$ and $a_{c12}$, whose estimation's accuracy slightly overpasses the objectives. The specification overrun appears respectively in 3\% and 1\% of the simulations. However, the global error budget after calibration complies by a large margin with the specification, which has been set to $2 \times 10^{-15} \, \mbox{ms}^{-2}$ (see table \ref{tab:Budget_cali}).

\begin{table}
\caption{Budget after calibration and correction of the EP measurement considering the simulated worst and mean estimation results.}
\label{tab:Budget_cali}
\begin{tabular}{lll}
\hline\noalign{\smallskip}
Disturbing term                                               & Contribution to the               & Contribution to the               \\
                                                              & EP test measurement               & EP test measurement               \\
                                                              & after calibration                 & after calibration                 \\
                                                              & and correction ($\mbox{ms}^{-2}$) & and correction ($\mbox{ms}^{-2}$) \\
                                                              & (worst estimation accuracy)       & (mean estimation accuracy)        \\
\noalign{\smallskip}\hline\noalign{\smallskip}
$a_{c11} \cdot T_{xx} \cdot \Delta_{x}$                               & $1.7 \times 10^{-16}$     & $4.2 \times 10^{-17}$             \\
$a_{c11} \cdot T_{xz} \cdot \Delta_{z}$                               & $2.1 \times 10^{-16}$     & $1.3 \times 10^{-16}$             \\
$a_{c11} \cdot T_{xy} \cdot \Delta_{y}$                               & $6.0 \times 10^{-18}$     & $1.5 \times 10^{-18}$             \\
$a_{c12} \cdot T_{yy} \cdot \Delta_{y} $                              & $3.5 \times 10^{-16}$     & $9.1 \times 10^{-17}$             \\
$a_{c13} \cdot T_{zz} \cdot \Delta_{z} $                              & $2.9 \times 10^{-16}$     & $8.2 \times 10^{-17}$             \\
$2 \cdot a_{d11} \cdot \Gamma_{res_{df},x}$                           & $1.0 \times 10^{-17}$     & $3.2 \times 10^{-18}$             \\
$2 \cdot a_{d12} \cdot \Gamma_{res_{df},y}$                           & $3.7 \times 10^{-17}$     & $2.4 \times 10^{-18}$             \\
$2 \cdot a_{d13} \cdot \Gamma_{res_{df},z}$                           & $7.5 \times 10^{-18}$     & $2.0 \times 10^{-18}$             \\
$4 \cdot K_{2,cxx} \cdot \Gamma_{app,dx} \cdot \Gamma_{res_{df},x}$   & $1.6 \times 10^{-17}$     & $3.5 \times 10^{-18}$             \\
$2 \cdot K_{2,dxx} \cdot (\Gamma_{app,dx}^2 + \Gamma_{res_{df},x}^2)$ & $7.1 \times 10^{-18}$     & $1.4 \times 10^{-18}$             \\
\noalign{\smallskip}\hline\noalign{\smallskip}
Total                                                                 & $1.1 \times 10^{-15}$     & $3.6 \times 10^{-16}$             \\
\noalign{\smallskip}\hline
\end{tabular}
\end{table}

\section{Conclusion}

For the success of the MICROSCOPE space experiment, a crucial problem is to reduce the impact of ten instrumental parameters that limit the measurement accuracy. Prior to any calibration of the instrument, the evaluated accuracy is not compatible with the EP test mission objective. To correct the measurement, an in-flight calibration is defined. Composed of ten in-orbit sessions, this calibration phase takes advantage of the satellite drag-free and pointing control, in addition to the possibility to move the masses inside the instrument core. Specific accelerations (linear, angular, amplitude, frequency and phase) can be generated thanks to the 6-D cold gas propulsion system. Other space fundamental physics missions such as STEP may have to follow this approach. The requirements for the satellite AOCS are similar to the ones necessary for the MICROSCOPE experiment (\cite{Sumner-STEP} and \cite{Overduin-STEP}).

A dedicated software is developed to simulate the proposed procedures. It includes a model of the drag-free loop, with the instrument, the satellite and the environment. The parameters estimations are extracted from the simulations with a simple data processing protocol based on a spectral analysis of the signals. The results comply with the mission specification, enabling the numerical validation of the foreseen calibration procedures.

The space EP test experiment will therefore be divided into different sequences including calibration sessions of the instrument and measurement sessions for the EP test, with the satellite either in inertial or in spinning pointing. Several calibration sessions are programmed in order to take into account the drift of the parameters values with time or mean temperature.

Furthermore, a software simulator dedicated to the sessions for the EP test is already available and the association of this tool with the calibration simulator will allow the simulation of the entire mission scenario and thus the preparation of the complete data processing protocol for the experiment.


\bibliographystyle{elsarticle-harv}
\bibliography{references}

\end{document}